\documentclass[journal]{IEEEtran} 
\IEEEoverridecommandlockouts
\usepackage{cite}
\usepackage{amsmath,amssymb,amsfonts}
\usepackage{algorithm,algorithmic}
\usepackage{graphicx}
\usepackage{textcomp}
\usepackage{xcolor}
\usepackage{subfigure}
\usepackage[utf8]{inputenc}
\usepackage[english]{babel}
\usepackage{graphicx}
\ifCLASSOPTIONcompsoc
    \usepackage[caption=false, font=normalsize, labelfont=sf, textfont=sf]{subfig}
\else
\usepackage[caption=false, font=footnotesize]{subfig}
\fi

\def\BibTeX{{\rm B\kern-.05em{\sc i\kern-.025em b}\kern-.08em
    T\kern-.1667em\lower.7ex\hbox{E}\kern-.125emX}}
\begin{document}

\title{A Screening Method for Power System Inertia Zones Identification\\
\thanks{This work has been supported by CRESYM projects BiGER Explore (https://cresym.eu/biger-explore/) and Harmony (https://cresym.eu/harmony/). Emails: $^1$\{R.Prasad, P.P.VergaraBarrios, A.Lekic\}@tudelft.nl, $^2$nppadhy@ee.iitr.ac.in, $^3$Robert.Dimitrovski@tennet.eu }
}
\author{\IEEEauthorblockN{Rashmi Prasad$^1$,
Pedro P. Vergara$^1$, Narayana Prasad Padhy$^2$, Robert Dimitrovski$^1$$^,$$^3$, and
Aleksandra Leki\'{c}$^1$}\\
\IEEEauthorblockA{$^1$Faculty of EEMCS, Delft University of Technology,
Delft, The Netherlands\\
$^2$Department of Electrical Engineering IIT Roorkee, India\\
$^3$TenneT GmbH, Bayreuth, Germany\\
}}
\maketitle

\IEEEpubidadjcol
\vspace{-3.5 mm}
\begin{abstract}
The heterogeneous distribution of frequency support from dispersed renewable generation sources results in varying inertia within the system. The effects of disturbances exhibit non-uniform variations contingent upon the disturbance's location and the affected region's topology and inertia. A screening method for inertia-zone identification is proposed considering the combination of network structure and generator inertia distribution that will aid in comprehending the response of nodes to disturbances. The nodes' dynamic nodal weight (DNW) is defined using maximal entropy random walk that defines each node's spreading power dynamics. Further, a modified weighted kmeans++ clustering technique is proposed using DNW to obtain the equivalent spatial points of each zone and the system to parameterize the inertia status of each zone. The impact of the proposed scheme is justified by simulating a modified IEEE 39 bus system with doubly-fed induction generator (DFIG) integration in the real-time digital simulator. 
\end{abstract}
\vspace{-2 mm}
\section{Introduction}
There has been a significant surge in renewable energy-based generation (REG) penetration to green the power industry \cite{murdock2021renewables}. The increasing penetration level of REG has promoted replacing conventional synchronous machine-based generation, thus introducing recurrent stability issues \cite{9360581}. Among various issues, the adverse effect of the decline in global inertia at distributed locations poses a great threat to stability, having a significant impact and frequent violation of the local frequency limit. Consequently, REG has participated in power-sharing and introduced technological advancements to support power system stability to counteract the stability issue \cite{chu2020towards}. 

The higher percentage of REG supporting the grid's frequency will form a dynamical change of the power grid due to the randomness in the REG generation source, which leads to variability in the reserve availability \cite{homan2021grid}. This results in spatiotemporal variations in power system inertia \cite{prasad2022spatiotemporal}. As system inertia becomes non-uniform across the network, some areas may experience varied levels of inertia deficiencies \cite{gu2020zonal,osbouei2019impact}. The transient fluctuations in local frequency cannot be captured by the temporal changes in the Center Of Inertia (COI), as the COI solely reflects the global frequency dynamics of the system. Consequently, there is a growing interest in studying zonal behavior to address frequency management within specific zones. The zoning of the network helps track the local responses for analysis, as it has become a crucial factor in determining the network's stability. 

The vulnerability status of the node is seen to be dependent on the unsymmetrical nature of the inertia distribution and on the locational and connectivity aspect of the network  \cite{misyris2023methodology}. Moreover, topology determines how fast the disturbances-induced energy dissipates to zero \cite{farmer2022evaluating}. The suggested method involves defining distinct inertia zones to monitor local inertia levels and assess the system's reaction to significant disruptions. These zones are defined as operational regions having similar inertial response characteristics. They are defined based on the inertia distributions of the generators contributing to the system dynamics and response to disturbances. Inertia zones can thus analyze the impact of changes in the distribution of generators and their inertia on the overall system stability, particularly in the context of renewable energy integration. They help assess how the response to disturbances varies across different grid areas.

Based on graph theory concepts, the Dynamic Nodal Weight (DNW) is proposed to estimate the spreading property of the dynamic behavior of the generator nodes in the network. Using the maximal entropy random walk centered on the maximum uncertainty principle provides the basis for DNW. This highlights the influence of network topology and inertia distribution across nodes and helps further find the effect on the disturbance of node dynamics. The modified weighted K-means++ scheme on the Fiedler eigenvectors of the Laplacian matrix of the dynamic graph and DNW is proposed for zoning. The network is clustered based on the inertia distribution features of the nodes and locates each network cluster's Spatial Equivalent Points (SEP). The DNW determines the weighted factor, combining network dynamics with the topology. The SEP provides a granular approach to spatial awareness of the area in response to the overall grid status, i.e., network working conditions and locational features. Also, the spatial equivalent distance is determined for the screening methodology to identify zones characterized by comparable inertial response characteristics flexibly. Identifying inertia zones needs regular updating since nodes might jump from one cluster to another in response to changing operating conditions and network topology. The algorithm provides easy and rapid generation of clusters for changing network topology and operating conditions due to its computation efficiency. 

\section{Methodology}
\begin{figure}
\centering
\subfigure{\includegraphics[height=4.5 cm,width=8.75cm]{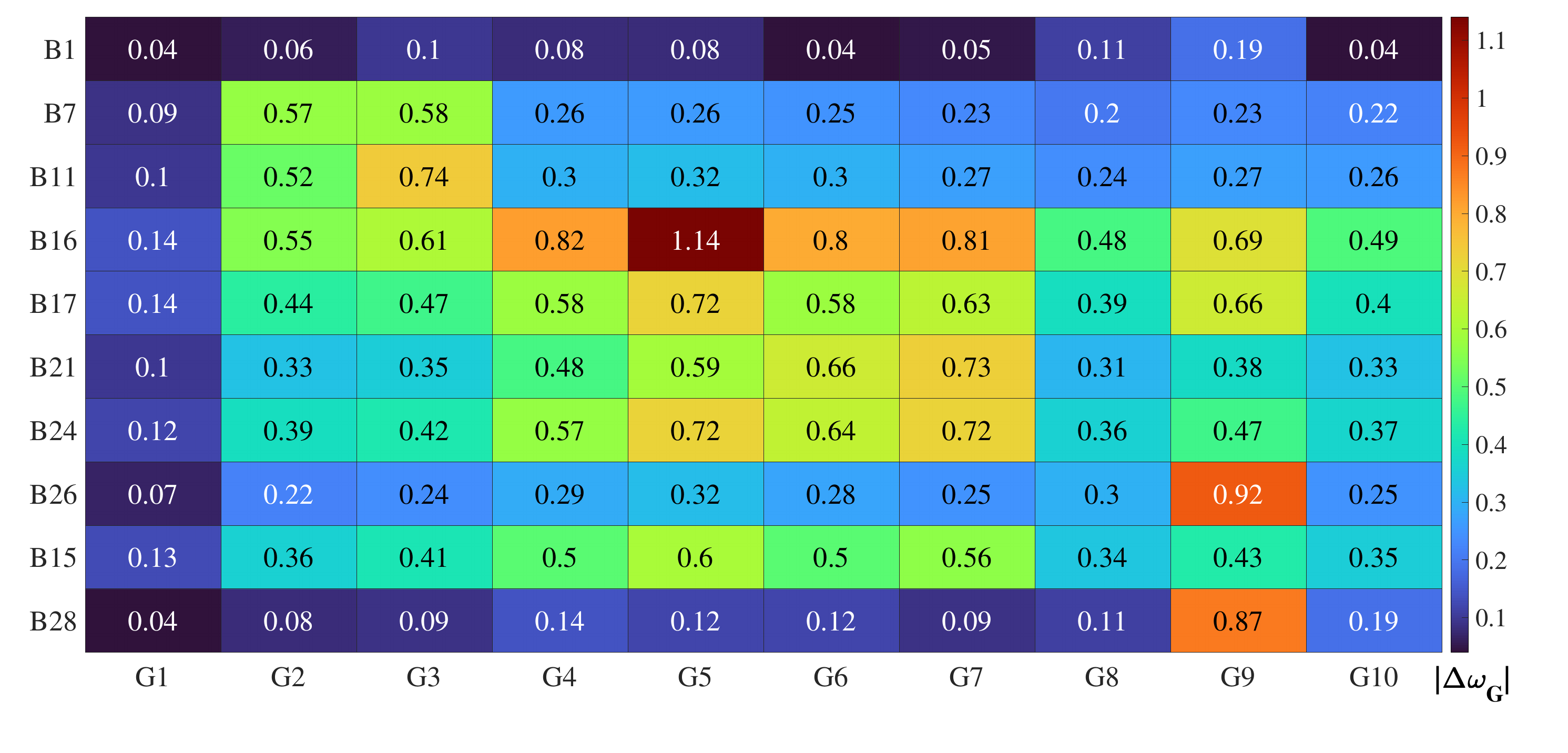}}
\caption{Case study shows the variation in rotor angle ($\Delta\omega_G$) of Generators numbered G1 to G10 when subjected to Fault at different buses (B\#).}
\label{fig_Case39}
\vspace{-5mm}
\end{figure}
The impact of disturbances varies non-uniformly depending on the disturbance's location and on the impacted area's topology and inertia. Case study in Fig. \ref{fig_Case39} shows the variation in rotor angle ($\Delta\omega_G$) of generators numbered G1 to G10 when subjected to similar duration and type fault at different load buses (B\#) of the standard IEEE 39 bus system \cite{athay1979practical}. It can be observed from Fig. \ref{fig_Case39}, that the location of faults triggers different levels of response of generators. Generator 1, irrespective of the location of the fault bus, maintains a uniform response, while most other generators seem highly dependent on the location of the fault. Moreover, the fault at load bus B1 also does not trigger the generator dynamics much. Thus, locational features highly influence the dynamic response.
\subsection{Grid Structure effect on Power System Dynamics}
Each node's dynamics are interconnected through the network structure, whereby the system topology unifies the dynamics of all nodes. Node stability is collectively defined by examining network attributes such as the degree and stability of surrounding nodes. Nodes with higher degrees indicate greater connectivity, enabling them to dissipate disturbance energy more effectively, resulting in increased support and reduced sensitivity to disturbances. The Spectral Graph Theory (SGT) provides a comprehensive network perspective by analyzing the Laplacian matrix's eigenpair. This eigenpair unravels some of the static and dynamic characteristics features of the electrical network \textit{i.e., spectrum and its associated eigenvectors} to explore a variety of network structural properties, including the degree of robustness. SGT relates the network's dynamic characteristics and topological information when the graph edges' weight is defined as the synchronizing power and nodes' weight by the inertia constant. The Laplacian matrix thus provides the relation of modal dynamics of linear oscillatory networks considering the generator (G) and load ($\kappa$) buses partition.  
The governing Differential-Algebraic Equation (DAE) of the system constituting the Laplacian matrix (L) is described as: 
\begin{gather}
\label{eq:swing}
 \frac{d}{dt}\begin{bmatrix} \Delta  \omega_{g} \\
 \Delta \omega_{\kappa}\end{bmatrix}
 =-
  \begin{bmatrix}
  M^{-1} & 0\\
  0 & 0 \end{bmatrix}
    \underbrace{\begin{bmatrix}
  P_{sGG} & 
P_{sG\kappa}\\
    P_{s\kappa G} &
     P_{s\kappa \kappa}
   \end{bmatrix}}_{L}
  \begin{bmatrix} \Delta \delta _{G} \\ \Delta \delta _{\kappa} \end{bmatrix}.
\end{gather}
Here $\Delta$ highlights a small change in value, 
${M=\operatorname{diag}(2H_i/(2\pi f_{n}))}$ in which $H_i$ is the inertia constant (in seconds) of the generator \textit{'i'}, $i\in G$, $f_n$ is the nominal frequency (Hz). Eq. (\ref{eq:swing}) is defined for the generator buses and the Wind Turbine Generator (WTG) contributing to frequency support. The above equation characterized the features of generator set \textit{$G$} with $N_g$ number of generators and the non-generator bus set $\kappa$ in a $N$ bus system, for a small change in active power \cite{ding2016graph}. The diagonal ($P_{sii}$) and off-diagonal ($P_{sij}$) elements of the synchronizing power coefficient matrix between nodes \textit{'i'} and \textit{'j'} in p.u. are given by:
\begin{equation} \label{eq:Psij}
P_{sij} = \left\{
\begin{array}{ll}
\sum\limits_{j=N_i}  V_i V_j B_{ij} cos( \delta_{i_o}-\delta_{j_o}), &  i=j, \\
- V_i V_j B_{ij} cos( \delta_{i_o}-\delta_{j_o}), & i\neq j,
\end{array} \right.
\end{equation}
$\forall\  i \in \{1\cdots N \}$ and $i^{th}$ bus neighbouring buses set is $N_i$. The Eq. (\ref{eq:Psij}) forms matrix L of (\ref{eq:swing}) which follows properties of the Laplacian matrix i.e. being symmetrical and semi-definite. For generator buses, voltages correspond to the pre-disturbance internal voltage (p.u.) of the generator with initial generator angle difference $\delta_{i_o}$. Moreover, for the load buses voltages are bus voltage. Susceptance term $B_{ij}$ of transfer admittance alone is considered while neglecting conductance. Subscript $\circ$ denotes the pre-disturbance value. The system dynamics can be sensed from the equation: 
\begin{equation}
    \frac{d\Delta \omega_g}{dt} =- \underbrace{M^{-1}[P_{sGG} - P_{sG\kappa}P_{s\kappa\kappa}^{-1}P_{s\kappa G}]}_{LM_{red}}\Delta \delta_{g}.
    \label{eq:delwg}
\end{equation}
For further analysis, the behavior of the state variables is obtained from the eigenvalue and eigenvector of $LM_{red}$. 
The Eq. (\ref{eq:delwg}) is valid for the wind turbine generator providing frequency support. The value of the initial inertia constant of the WTG-based DFIG can be obtained \cite{liu2020coherency}. 
\subsection{Dynamic Nodal Weight}
The DNW estimates the spreading property of the dynamical behavior of the generator nodes in the network. In other words, it represents the importance of individual nodes in their dynamic behavior. The DNW represents the inertia distribution, which characterizes the trend of diffusion of generator inertia effect to non-generator nodes and its influence on inter-generator interactions. 

DNW uses the Maximal Entropy Random Walk (MERW), a mathematical model, to calculate the DNW. The traditional random walk has already been used to calculate the nodal inertia index using the eigenvector centrality (EVC) \cite{prasad2022spatiotemporal}. Eigenvector centrality emphasizes its relevance as a mathematical model of the information flow process that can reveal positional biases that impact the distribution of ideas within networks \cite{bienenstock2022eigenvector}. However, the general random walk violates the Maximum Uncertainty Principle (MUP), leading to a biased prediction. The MERW algorithm can adequately implement the MUP in a weighted graph and obtain an equilibrium distribution with strong localization properties \cite{yu2013maximal}.
As the partitioned matrix of the generator and non-generator nodes is considered, the first step is to calculate the DNW of the generator nodes. U and $\Lambda$ are the eigenvectors and eigenvalue matrices of the $LM_{red}$ matrix. 
\begin{algorithm}[H]
 \caption{Dynamic Nodal Weight}
\begin{algorithmic}[1]
 \STATE Pre-requisite: $LM_{red}$: Non-negative 
 \STATE $LM_{red}$*U = U*$\Lambda$ 
 \STATE $\lambda_m$=max($\Lambda$) 
 \STATE $u_m$=$U|_{\lambda_m}$,  $u_m \in \mathbf{R}^{N_g*1}$ 
 \STATE Transition Probability Matrix: $P_{M_{ij}}$ = $\frac{LM_{{red}_{ij}}u_{m_j}}{\lambda_m u_{m_i}}$ 
 \STATE Stationary Distribution of MERW: $\Pi_{M_i}\ =\ u_{m_i}^2$
 \end{algorithmic}
 \label{MERW}
\end{algorithm}
\vspace{-3 mm}
Algorithm \ref{MERW} calculates $\Pi_M$, which is the DNW of the generator nodes. The transition probability matrix ($P_{M}$) needs to be non-negative as the Frobenius-Perron Theorem guarantees maximum eigenvalue ($\lambda_m$) and its corresponding eigenvector $u_m$ to be positive.  
\begin{figure}
\centering
\subfigure{\includegraphics[height=2.8 cm,width=3.5 cm]{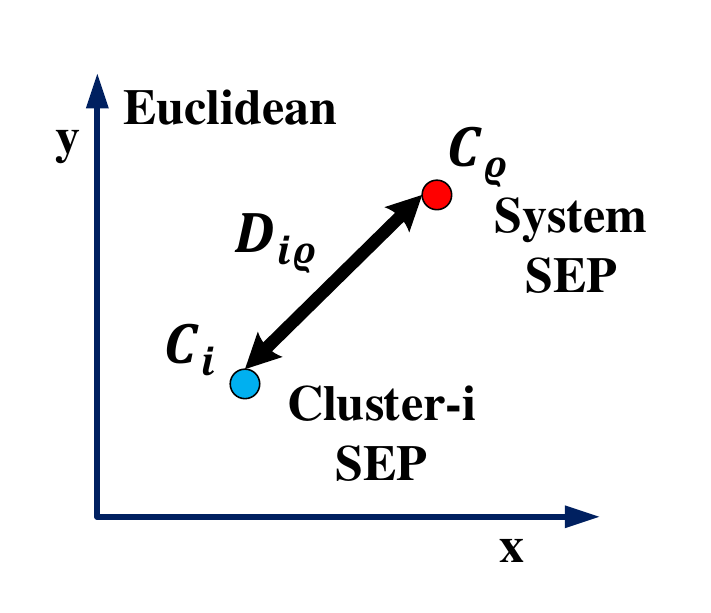}}
\caption{Distance between System and each cluster SEP, here i $\in \{1, ..., k\}$. }
\label{fig_Distance}
\vspace{-6 mm}
\end{figure}
For obtaining the DNW of the non-generator bus, the effect of the neighboring influence of the generator nodes is considered by calculating load only path of reaching the nodes. It can be found from (\ref{eq:swing}) when expressed in the form of eigenvalue and eigenvectors ${(-L\upsilon=\lambda M \upsilon )}$ \cite{ding2016graph}, expanding the relation, $P_{s\kappa G} \upsilon_G + P_{s\kappa\kappa}\upsilon_{\kappa} = 0$, where the $\upsilon_G$ and $\upsilon_{\kappa}$ are the partitioned generator and non-generating node eigenvector, thus: 
    $\upsilon_{\kappa}=-P_{s\kappa\kappa}^{-1}P_{s\kappa G}\upsilon_G.$
Thus, the eigenvector of the non-generating nodes is dependent on the structural distribution characteristics of the generating node eigenvector. The level of influence directly depends on the path that different generators have to load nodes without encountering any other generator node. So, net nodal weight is obtained by:
\begin{equation}
    \Pi_{M_{net}}=[\Pi_M ; -(P_{s\kappa\kappa}^{-1}P_{s\kappa G})\Pi_M ].
\end{equation}
\subsection{Inertia Zone and Spatial Equivalence}
A dynamic similarity is observed among a group of nodes within a power network, as defined by specific graph criteria. Zoning imparts analogous behavior to these node clusters, with the cluster center playing a vital role in operational and control aspects. Furthermore, the spatial structure of the network significantly impacts frequency dynamics, leading to the utilization of graph spectral properties for clustering networks with comparable inertia characteristics. The method uses data points: i) eigenvectors corresponding to the \textit{'r'} slowest mode of the $LM_{red}$ matrix of the dynamic graph, ii) DNW considers the nodal features to form a multidimensional feature data vector. The improved weighted kmeans++ Algorithm \ref{Kmeans} modifies the network by introducing the weighted mean, which is additionally influenced by the DNW input. 
\begin{algorithm}[H]
 \caption{Modified Weighted kmeans++ Algorithm}
\begin{algorithmic}[1]
\REQUIRE \textbf{Defining Initial centroid and number of clusters}

 \STATE Select the first centroid ($C_o$) at random from the data points. 
 \STATE Assign the centroid ($C_o$) to the cluster set {$C_{oi}$}.
 \STATE Calculate distance ($d_i$) of all data points from ($C_{oi}$).
 \STATE Calculate the intracluster distance ($S_i$)= data points distance to nearest centroid.
 \STATE Data Point corresponding to max distance is assigned as next centroid ($C_o$).
 \STATE Repeat 2-5 till $S_i$ and $S_{i-1}$ comes very near.

 \REQUIRE \textbf{Inertia Zone:} Suppose the stopping criteria is fulfilled at $k^{th}$ step, where $S_k$ $\approx$ $S_{k-1}$. So the number of zones is k and the initial centroid= $C_{oi}$, where i=k.
 \STATE Assign points to k different clusters based on the minimum distance between the centroid and data point
\STATE Recompute Centroid ($C_k$) using the weighted mean taking weight (w) = (1./$\Pi_{M_{net}}$), 

$W=\frac{\sum\limits_{i=1}^{n} w_i*x_i}{\sum\limits_{i=1}^{n} w_i}$

\STATE Repeat 7-8 while checking the minimizing criteria of intracluster distance.
 \end{algorithmic}
 \label{Kmeans}
\end{algorithm}
After satisfying the stopping criteria, the centroid ($C_{k}$) of each cluster is obtained and called the spatial equivalent point (SEP) to get the overall centroid ($C_{\varrho}$) using the weighted mean. The nodal weight of each cluster is summed to get the cluster nodal weight, i.e., system SEP. The centroid ($C_{k}$) of each cluster represents the group's characteristics and behavior based on the network's spatial awareness. 

In a traditional power system where synchronous generators are placed at a remote location, higher utilization of the inertia occurs at a lower degree node, typically peripheral nodes in a power system network. However, in a low-inertia power system, the optimal use of inertial support units is to distribute inertia throughout the network to slow down disturbance propagation that can benefit frequency transient stability \cite{farmer2022evaluating}. Thus, REG has significant impacts depending on the varied locations of the power system. So, accordingly, the support mechanism needs to be defined for REG located at different locations. The algorithm delivers zoning to identify the sub-area for more localized monitoring. Further, the authors employ DNW and SEP matrices, which quantify the nodes' relative robustness in the network by considering spatial considerations.

To find the Spatial Equivalent Euclidean Distance (SED) of the cluster SEP ($C_i, i\in \{1, ..., k\}$) from the system SEP ($C_\varrho$) as shown in Fig. \ref{fig_Distance} and calculated as: 
\begin{equation}    
SED(i)=\frac{D_{i\varrho}}{\max(D_{i\varrho})}, \qquad i\in \{1, ..., k\}.
\label{eq:SED}
\end{equation}
The SED provides the idea of the status of equivalent inertia in the zones. Higher SED corresponds to zones of lower inertia distribution. This helps in parameterizing the inertia status of the zones. 
\section{Result and Discussions}
In this section, we have deliberated upon the findings from four test scenarios within the IEEE 39 bus system. These test case scenarios are executed in the Real-Time Digital Simulator (RTDS), placing DFIG-based WTGs at various system buses. The computational work is performed in the MATLAB environment. The information exchange between RTDS and MATLAB is done through a TCP/IP connection \cite{prasad2019data}. The modified discrete Fourier transform analysis is performed on the voltage waveform to obtain the RMS voltage and angle. 

A generation mix of synchronous generator and DFIG-based WTG supply 6097.1 MW load of IEEE 39 bus system, as per the scenario discussed:  
\begin{itemize}
    \item Scenario 1: Base Case with 10 Synchronous Generators (SG).
    \item Scenario 2:  SG is replaced by DFIG-WTG at buses 30 and 33, providing an inertia constant of 5.6 each. 
    \item Scenario 3: In the scenario 2 system, an additional DFIG-WTG is installed at bus 28, rated at 255 MW and possessing an inertia constant of 5.6 s. The equivalent load is evenly distributed.
    \item Scenario 4:  In the scenario 2 system, an additional DFIG-WTG is installed at bus 19, rated at 255 MW and possessing an inertia constant of 5.6 s. The equivalent load is evenly distributed.
\end{itemize}
 \begin{figure}
\centering
\subfigure[]{\includegraphics[height=5.0 cm,width=4.37 cm]{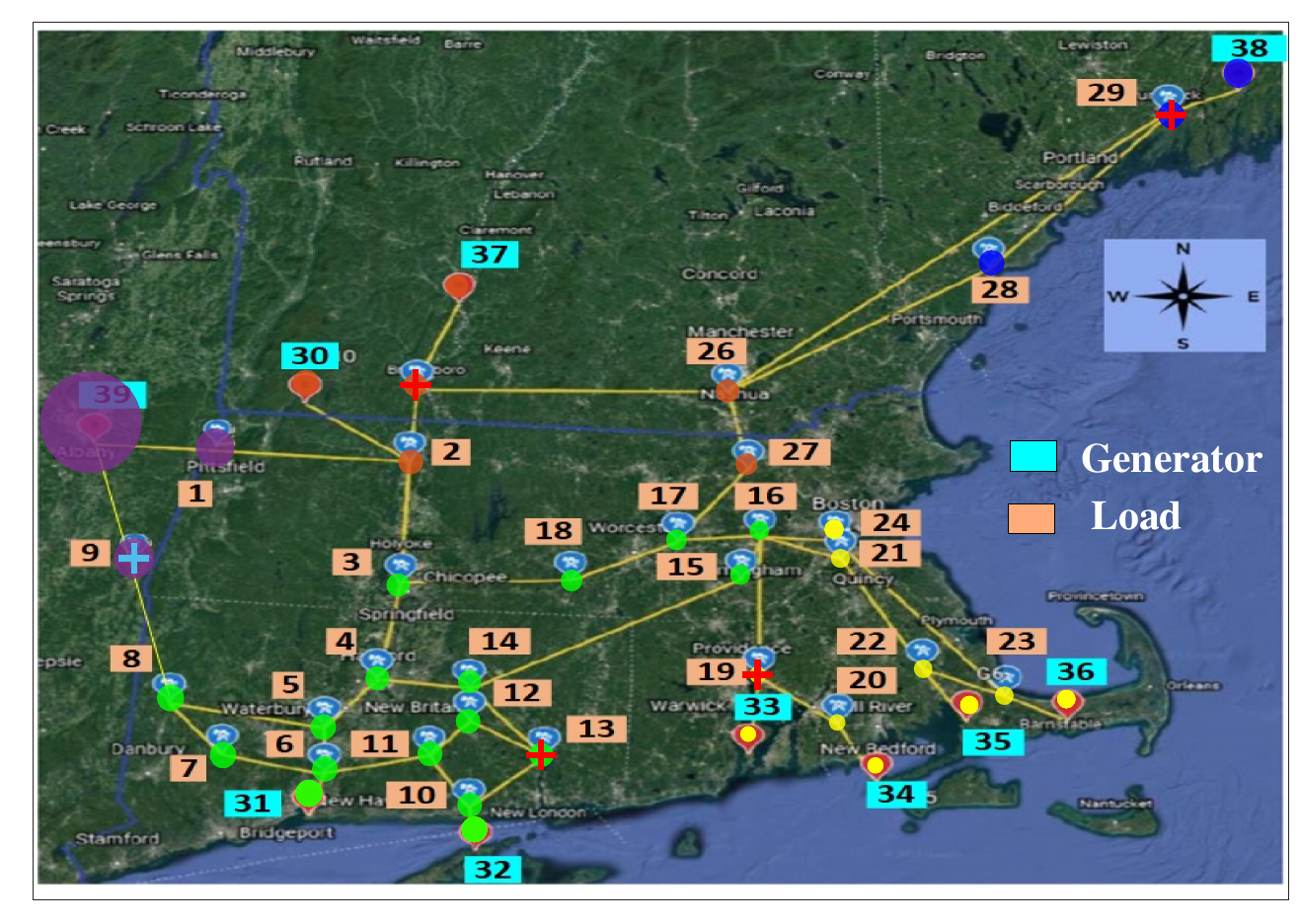}}
\subfigure[]{\includegraphics[height=5.0 cm,width=4.37 cm]{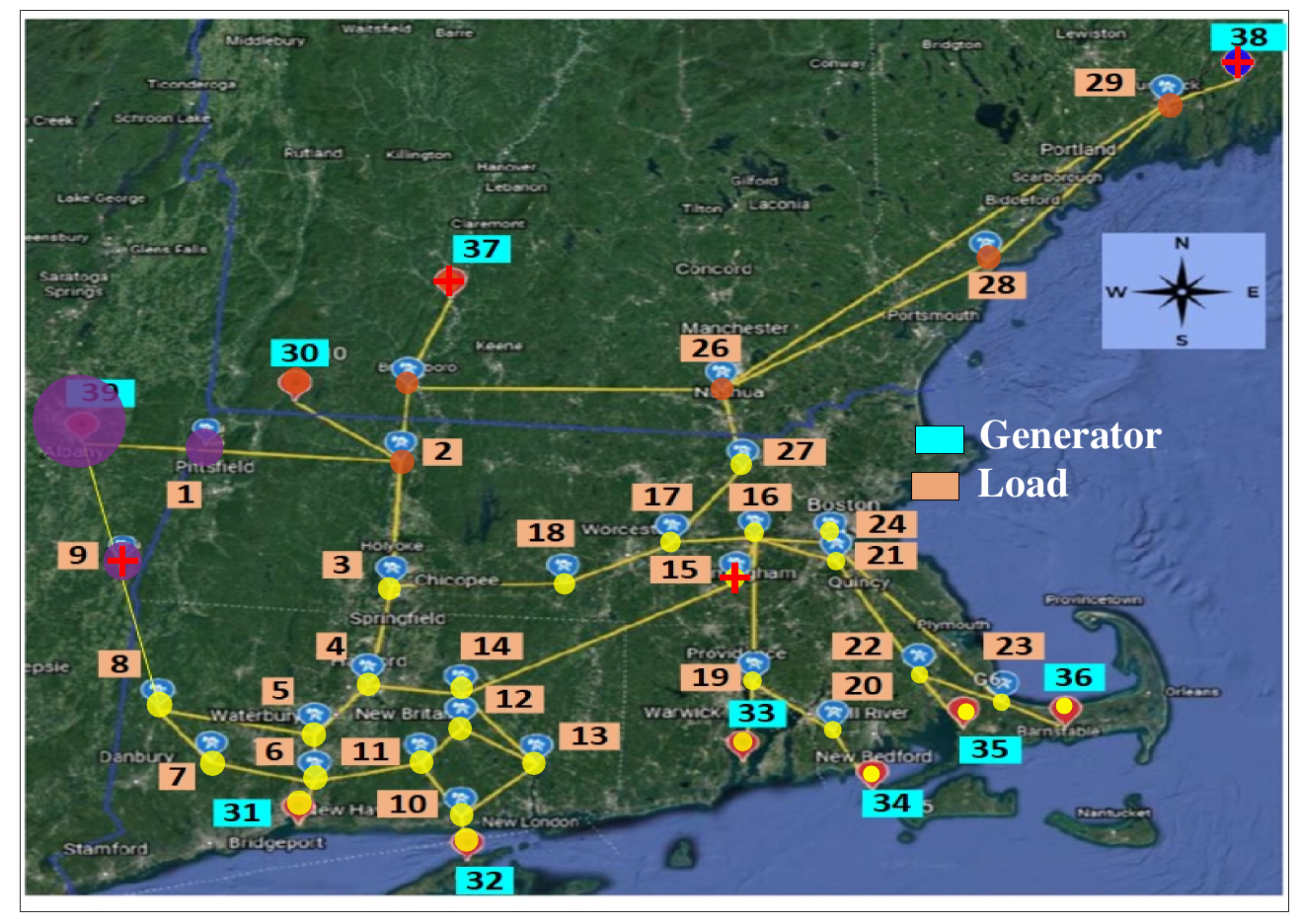}}
\caption{ Inertia Zones Shown by various color dots and cluster SEP marked as Red Cross for (a) Scenario 1;  (b) Scenario 2.}
\label{fig_Scenario12}
\vspace{-3 mm}
\end{figure}
\begin{figure}
\centering
\subfigure{\includegraphics[height=4.8 cm,width=8.5cm]{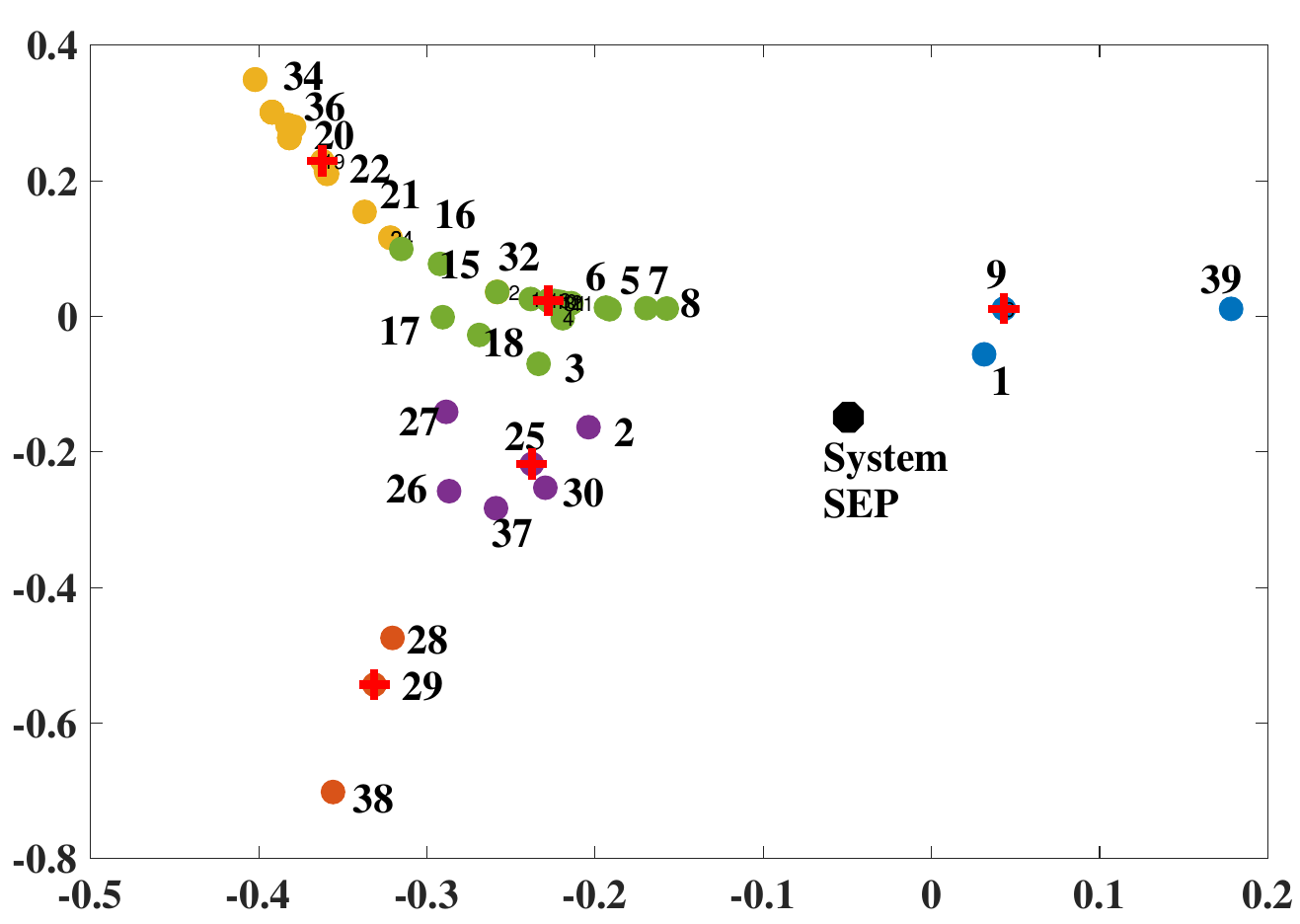}}
\caption{Plot between eigenvector corresponding to two slowest modes indicating system SEP by a black dot and cluster SEP by a red cross.}
\label{fig_Eigengraph}
\vspace{-3 mm}
\end{figure}

Fig. \ref{fig_Scenario12} illustrates replacing the SG with DFIG-WTG, providing a higher value of inertia constant leads to less number clusters. It can be marked from the figure that there is a more uniform inertia distribution in Scenario 2 with the radius of the circle displaying inertia distribution as obtained by ($1/\Pi_{M_{net}, i}$ for $i$ representing rows of vector $\Pi_{M_{net}}$).

Fig. \ref{fig_Eigengraph} illustrates the plot depicting the eigenvector associated with the two slowest modes for scenario 1. Each cluster SEP is identified in the figure by a red cross, serving as a reference node for frequency measurement within the respective cluster. Measuring the frequency at this reference node provides the frequency response for all nodes in that cluster. The system SEP is computed using a weighted mean, offering insights into inertia zones both near and distant from the system's SEP. This calculation reveals a lower inertia distribution in regions farther from the system's SEP, while the system SEP itself is situated closer to the SEP zone having a higher inertia distribution.

\begin{figure}
\centering
\subfigure[]{\includegraphics[height=5.0 cm,width=4.37 cm]{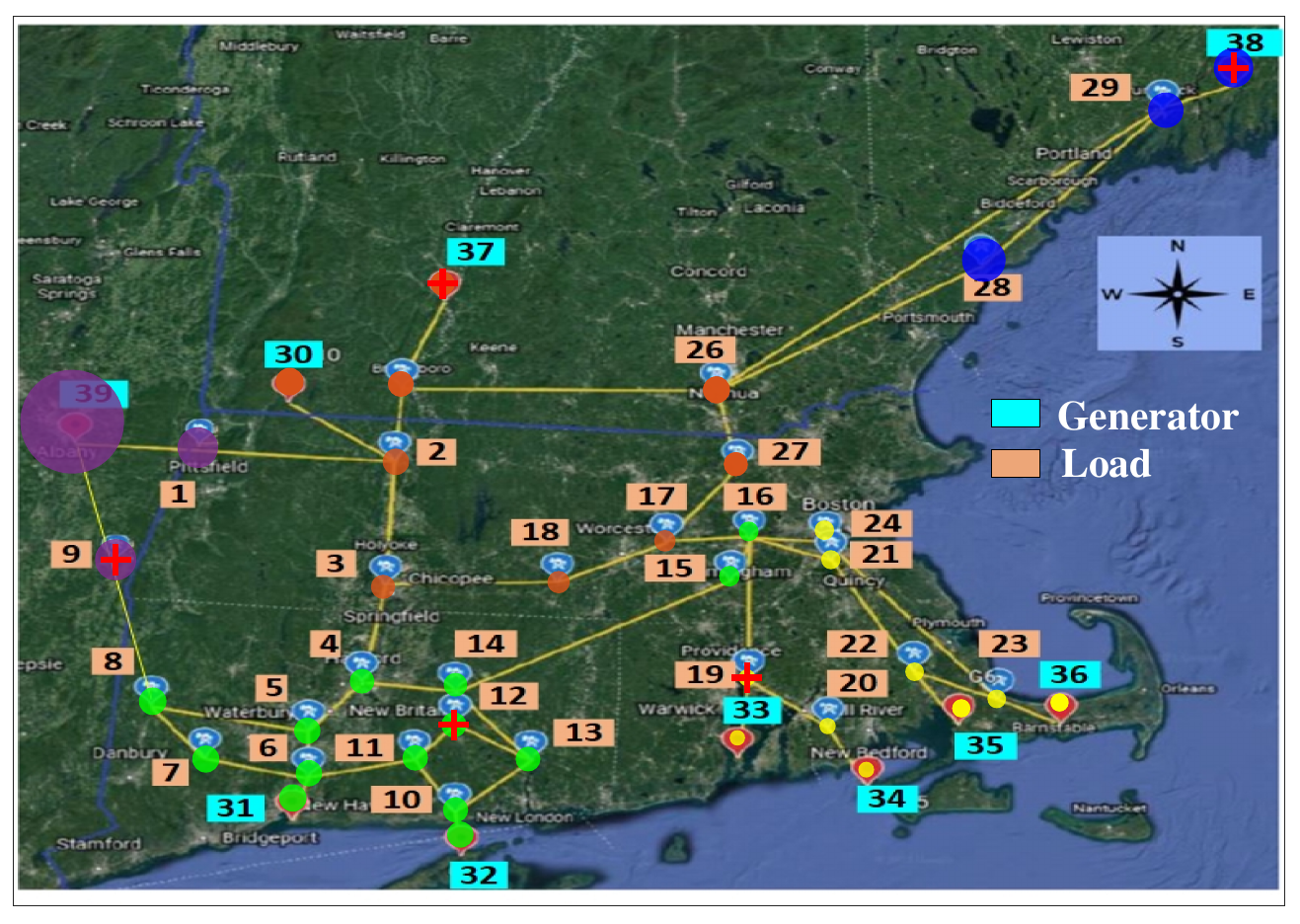}}
\subfigure[]{\includegraphics[height=5.0 cm,width=4.37 cm]{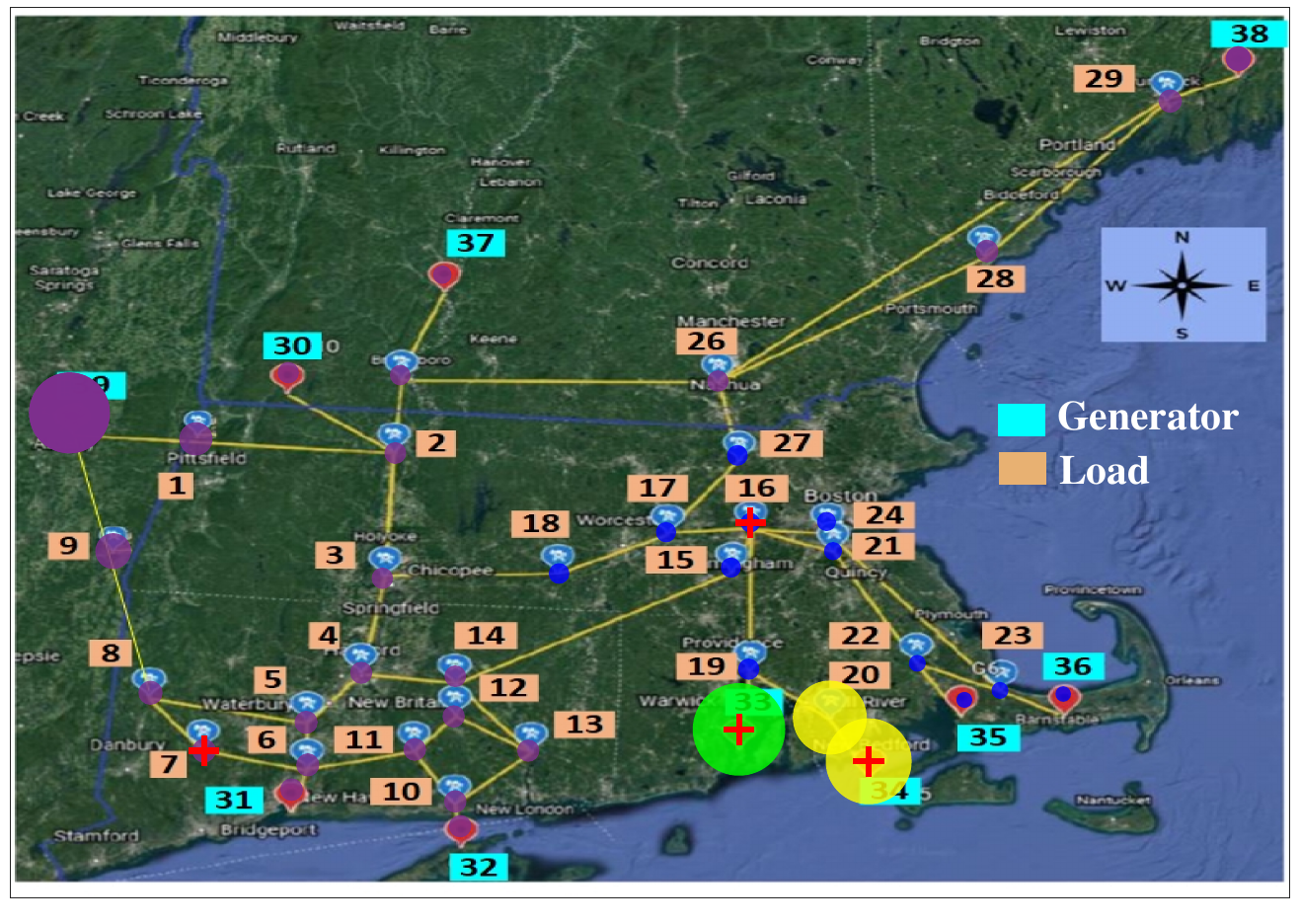}}
\caption{ Inertia Zones Shown by various color dots and cluster SEP marked as a red cross for (a) Scenario 3;  (b) Scenario 4.}
\label{fig_Scenario34}
\vspace{-5 mm}
\end{figure}

\begin{figure}
\centering
\subfigure{\includegraphics[height=4 cm,width=8 cm]{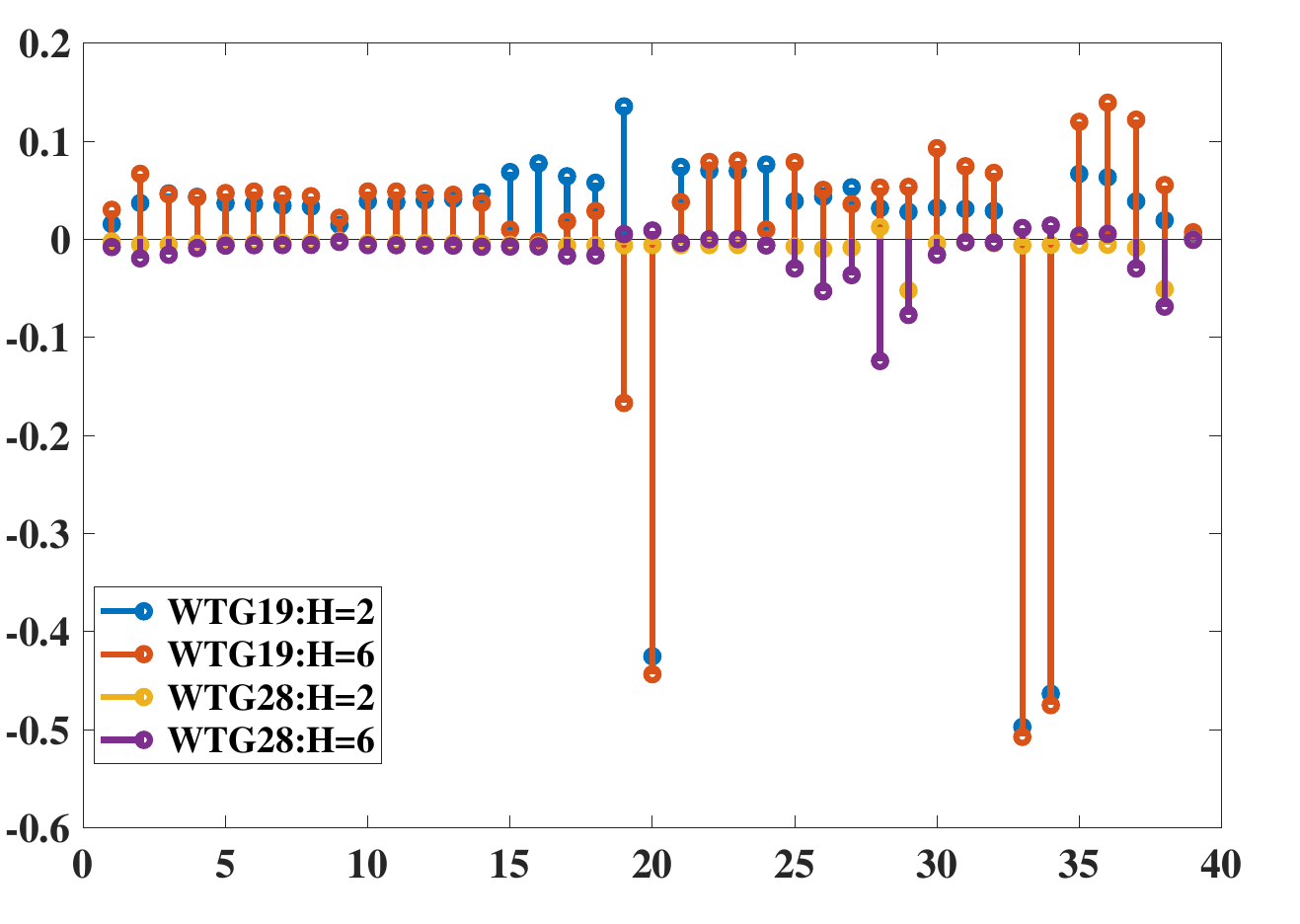}}
\caption{Variation in Inertia distribution in Scenario 3 and 4 when inertia in WTG varies from 2 to 6 seconds.}
\label{fig_DNW39}
\vspace{-5 mm}
\end{figure}

Intentionally, two locations of WTG are taken, one with a relatively higher DNW (Scenario 4) and one with a low DNW (Scenario 3). From Fig. \ref{fig_Scenario34}, an observation is made that the path of the release of generator energy, if blocked by another generator, will only contribute to the variation of direct path nodes. An essential aspect of the study is the variation in nodes 33, 34, and 20 is relatively higher than the standard test system value when WTG is placed at bus 19. The WTG at node 19 does not provide a non-generator path to the rest of the network from nodes 34 and 35. So none of the non-generator nodes except node 20 are influenced by the generator node at 34 and 35 and are solely influenced by the generator factor of node 19. The second observation is from the size of dots reduced, \textit{i.e.} the overall variation of the nodes increases with the WTG placed at bus 19 than 28 in the system. Fig. \ref{fig_DNW39} justify the observation of Fig. \ref{fig_Case39} \textit{i.e.} the variation in nodes with lower inertia distribution results in more variation in response. WTG placed at bus\#19 varies its inertia from 2 to 6 seconds, then the variation in inertia distribution is varied more when the inertia constant of WTG placed at bus 28 varies. 
 Also, it can be seen that the number of zones is determined by Algorithm 2, thus Fig.\ref{fig_Scenario12} and \ref{fig_Scenario34} have different numbers of zones formed.

\textbf{Sensitivity Analysis of DNW}: By using the matrix perturbation theory using the analytic expansion of eigenvalues and eigenvectors as matrix equations, namely the Sylvester equation, the variation in the eigenvector is studied \cite{bamieh2020tutorial}. 
As mentioned above, the DNW is reflected from the eigenvector ($u_m$ ) corresponding to the matrix largest eigenvalue $\lambda_m$. So the variation in $\upsilon_m$ is marked:

The initial matrix operation follows:
\begin{align*} 
LM_{{red}_o} * U_o &=  U_o* \Lambda_o \\ 
W^*_o * LM_{{red}_o} &=\Lambda_o W^*_o 
\end{align*}
where $U_o$ and $W^*_o$ are the right and left eigenvector respectively and $\Lambda_o$ is the eigenvalue of matrix $LM_{{red}_o}$. Perturbation Expansion of matrix $LM_{{red}_o}$, for some small parameter $\epsilon$ is defined as:$LM_{{red}_\epsilon}=LM_{{red}_o} + \epsilon LM_{{red}_1}$, expanding:
\begin{multline*}
(LM_{{red}_o} + \epsilon LM_{{red}_1})(U_o + \epsilon U_1 +\cdots) = (U_o + \epsilon U_1 +\cdots) \\
(\Lambda_o + \epsilon \Lambda_1 +\cdots)
\vspace{-1 mm}
\end{multline*}
Variations are obtained as:
\begin{align*} 
\Lambda_1 &= diagonal(W^*_oLM_{{red}_1}U_o) \\ 
U_1 &= -U_o(\Upsilon^{+o}o(W^*_oLM_{{red}_1}U_o))
\vspace{-1 mm}
\end{align*}
where, $\Upsilon^{+o}=\frac{1}{\lambda_{oi}-\lambda_{oj}}$ if $i\neq j$, $\Upsilon^{+o}=0$, if i=j. 

Analysis using the IEEE 39 bus system: Variation in voltage, angle, and inertia is varied by 0.2 units of the initial value. $U_{o-fd}$ eigenvector corresponding to matrix's ($LM_{{red}_o}$) largest eigenvalue. 
$U_{1-fd}$ is variation in largest eigenvector.

 $U_{1var}$=sum(abs(($U_{1-fd} ./U_{o-fd}$))
\begin{table}[htbp]
\caption{Sensitivity Analysis of DNW}
\begin{center}
\begin{tabular}{|c|c|}
\hline
\cline{1-2}
{Variation in eigenvector with variation of voltage} & {  } \\
{$V_{1var-\Delta E}$} & {0.0301}  \\
\hline
{Variation in eigenvector with variation of voltage angle} & {  }\\
{$V_{1var-\Delta \delta}$-} & {0.0104}\\
\hline
\cline{1-2}
{Variation in eigenvector with variation of inertia} & {  } \\
$V_{1var-\Delta H}$&0.2044 \\
\hline

\end{tabular}
\label{tab1}
\end{center}
\vspace{-3 mm}
\end{table}
It has been seen that the impact of voltage variations and the exciter elements are less in compassion to variation in inertia constant changes, the inertia distribution regions are just slightly modified. 
The inertia distribution is sensitive towards grid topology change or inertia of generator change. The simulation helps us understand the importance of monitoring the inertia variation from the WTG and its impact on the system. The study of zoning will thus prepare the system for the upcoming varying inertia scenario with higher penetration of REG. 
\section{Conclusion}
The frequency of the nodes significantly governs the input to the frequency support loop of the WTG. The non-uniform distribution of inertia is noted in conjunction with the diverse frequency support provided by dispersed REG. Consequently, specific groups of nodes exhibiting similar response patterns are zoned to enhance the localized analysis of the system. The dynamic nodal weight and system spatial equivalent distance, which analyze the grid operating status, are calculated for this. The article enhances the WTG inertial frequency support state-of-the-art by considering the spatial effect of the grid stability criteria. DNW represents the spatio-temporal dynamics of the network. Using the maximal entropy random walk technique finds the DNW accurately, enhances the algorithm's robustness, and is also computationally faster with 0.003085 s. The detailed model and control strategy are implemented in the real-time digital simulator, which provides a realistic response from the WTG, even in the case of a dynamic response. The technique reduces the overall computational complexity in providing the local responses considering the network features. 

\bibliographystyle{IEEEtran}
\bibliography{reference}
\end{document}